\documentclass[1p]{elsarticle}

\def\t{\times}
\usepackage{amssymb}
\usepackage{here}
\usepackage{graphicx}
\usepackage{color}
\usepackage{lineno}

\newcommand{\AmS}{{\protect\the\textfont2 A\kern-.1667em\lower.5ex\hbox{M}\kern-.125emS}}
\journal{Nuclear Instruments and Methods A}

\begin{document}

\begin{frontmatter}


\title{Lifetime-Extended MCP-PMT}



\author[A]{T.~Jinno}
\author[A]{T.~Mori}
\ead{mori@hepl.phys.nagoya-u.ac.jp}
\author[A]{T.~Ohshima}
\author[A]{Y.~Arita}
\author[A]{K.~Inami}
\author[B]{T.~Ihara}
\author[B]{H.~Nishizawa}
\author[B]{and T.~Sasaki}


\address[A]{Department of Physics, Nagoya University, Chikusa, Nagoya 464-8602, Japan}
\address[B]{Hamamatsu Photonics K.K. Electron Tube Division Manuf.$\#1$, Dept.43, 314-5 Shimokanzo, Iwata, Shizuoka 438-0193, Japan}

\begin{abstract}
In order to develop a long-lifetime MCP-PMT under high rates of circumstance, 
we investigated the degradation of the quantum efficiency ($QE$) of PMT's with 
a multialkali photocathode. 
We found that not only positive ions, but also neutral residual gases 
would damage the photocathode resulting in an enhancement of the work function;
their countermeasures were established in newly manufactured square-shaped 
MCP-PMT's with 4 or 4$\times$4 multi-anodes. 
The performances of the PMT's were measured: $QE$ was stable up to an integrated
amount of anode output charge of $2-3$ C/cm$^2$, while keeping other basic 
performances steady, such as the time resolution for single photons 
($\sigma_{\rm TTS}$) of $\simeq40$ ps, 
a photoelectron collection efficiency ($CE$) of 60\%, 
a multiplication gain ($G$) of a few $\times 10^{6}$, 
and dark counts of $20-300$ Hz. 
The causes of $QE$ degradation are discussed. 
\vspace{1pc}
\end{abstract}


\end{frontmatter}


\section{Introduction}



The most difficult R\&D item of a microchannel plate (MCP) 
photo-multiplier tube (PMT) is to settle the subject for protecting 
the photocathode from its $QE$ degradation under a high counting rate 
for a long experimental period. 

We previously found \cite{Kishi, Inami} that 
an Aluminum prevention layer from ion-feedbacks plays an essential role 
for this problem, and an MCP-PMT (Hamamatsu Photonics K.K. (HPK) R3809U-50-11X)
with a multialkali photocathode kept its performance up to 
the integrated amount of 
output charge of $Q >$ 2.6 C/cm$^2$, corresponding to more than a 14-year 
time duration \cite{ZERO} under our supposed Super-KEKB/Belle-II 
\cite{superB}. 

Our development of MCP-PMT's for the practical use on a newly proposed K/$\pi$ 
particle identification detector, a time-of-propagation (TOP) counter 
\cite{paper2}, was then followed, 
in order to enlarge its effective area, change its cross section from circular 
to square-shaped and to equip it with multi-anodes.
The characteristics and performance of the developed MCP-PMT, SL10, are 
satisfactory, as reported in \cite{Inami}.
However, unexpected $QE$ degradation, which brings the above work to naught,
was encountered after the previous publication.

While studying photocathodes, such as bi-alkali, multialkali, and GaAs(P) 
materials, including a negative electron affinity (NEA) property, has in general
long-continued, their details and certain understandings have not been
established due to complexity from a solid-state physics viewpoint.  
Especially investigations of the $QE$ deterioration
of photocathodes in the case of MCP-PMT are scarce.
We carried out an exhaustive study on this phenomenon, by manufacturing 
about 30 different versions of MCP-PMT's. The resulting SL10 with a 
${\rm Na_2KSb(Cs)}$ multialkali photocathode survived up to 
$Q=2-3$ C/cm$^2$.
It could satisfactory perform well over 10 equivalent-years 
at the Super-KEKB/Belle-II. 
New insights were gained, about which we would like to report here.  

We hereafter refer to MCP-PMT as simply PMT, for brevity. 

\section{PMT and Setups}

\subsection{SL10}

While the basic characteristics of the SL10 can be found in Table~1 
in \cite{Inami}, slight variations were made for individual SL10 
versions in this research. 
Table~\ref{T-1} is presented for the reader's convenience. 
%
\begin{table*}[thb]
\begin{center}
\caption{\small Basic characteristics of SL10. 
Some variations can be found, compared to Table~1 in \cite{Inami}. }
\label{T-1}
\small
\begin{tabular}{|c|c|} \hline\hline 
\hspace*{18 mm} items \hspace*{18 mm} & 
\hspace*{18 mm} SL10 \hspace*{18 mm} \\ \hline\hline 
photocathode & multialkali ${\rm Na_2KSb(Cs)}$ \\ \hline\hline
window & borosilicate glass \\
effective area & 22$\times$22 (mm$^2$) \\
QE @ $\lambda=400$ nm & $\sim 20$ (\%) \\
channel-diameter of MCP & 10 ($\mu$m) \\
length-to-diameter ratio of MCP & 40 \\
MCP aperture & $\sim$ 60 (\%) \\
bias angle & 13($^o$) \\ \hline 
Al prevention layer~$^{\sharp}$ & ON \\
anodes & 4, ~4$\times$4 \\ \hline\hline
 & gap (mm) / voltage (kV) $^{\flat}$ \\  \hline
photocathode - 1st MCP & 2.0 ~/~ ~0.2  \\ 
1st MCP in - out             & 0.4 ~/~ ~1.0  \\
1st - 2nd MCP's        & 1.0 ~/~ ~1.0 \\
2nd MCP in - out              & 0.4 ~/~ ~1.0  \\
2nd MCP - anodes       & 1.0 ~/~ ~0.6  \\ \hline\hline
HV supplied & 3.4-3.8 (kV) \\
gain & $(1-3)\times 10^6$ \\
$\sigma_{\rm TTS}$ & 30-40 (ps) \\
dark counts & ${\cal O}(10-10^4)$ (Hz) \\
$CE$ & 60 (\%) \\ \hline\hline
\end{tabular}
\normalsize
\end{center}
{\small $\sharp$ For the effect of Al prevention layer, 
see the text and Table~2.}

\noindent
{\small $\flat$ Placing the Al prevention layer on the 2nd MCP yields a distance
of 1 mm between MCP's.} 
\end{table*}

\subsection{Setup for a $QE$ degradation measurement} 

The $QE$ degradation of PMT's is measured in terms of an integrated amount 
of output charge, $\sum_{\rm Q}$, by the system, illustrated in Fig.1 
in~\cite{Kishi}. 
The system consists of an LED, a light pulser, a filter, a standard PMT 
for calibration, and PMT's tested, in a black box.  
An LED ($\lambda\simeq 400$ nm) was pulsed (10 ns-wide) at repetition 
rates of $f=1-20$ kHz, which yielded the number of observed 
photoelectrons, $N_{\rm p.e.}=20-50$ per pulse, corresponding to 
the output charge from an anode of $\sum_{\rm Q}=2-10$ mC/cm$^2$/day 
under an expected Belle-II condition, $\sum_{\rm Q}$=0.16 C/cm$^2$ per year.
During continuous irradiation, the performance of SL10's for single photons 
was monitored every 1-3 days, using a light pulser PLP ($\lambda=408$ nm 
with a duration of 50 ps (FWHM) and a jitter of $\pm 10$ ps). 

\subsection{Setup for a $QE$ $\lambda$-dependence measurement}

To measure the $QE(\lambda)$-spectrum, a monochromator system was prepared 
by a Halogen lamp, a monochromator ($\lambda=350-900$ nm) with 
a wavelength-cut filter, a Si photodiode as a standard photo-device, and 
a pico-ampere meter to measure the photo-currents, as illustrated in Fig.5 
in \cite{Kishi}.  

\section{Al Prevention Layer}

It was concluded in our previous study that the $QE$ deterioration of 
the photocathode could be attributed to positive ion-feedbacks, produced 
in the multiplication process of secondary electrons. 
The bias-angle setting of the MCP might not be powerful enough as a single 
measure of ion-feedbacks.  
We succeeded to protect the photocathode from ion-feedbacks by 
equipping a thin Aluminum layer between the photocathode and 
the 1st MCP-layer. 
The effect of the Al-layer was proved upon removing the after-pulses 
due to ion-feedbacks, and the $QE$ of the photocathode at 400 nm
for a PMT R3809U-50-11X 
had remained stable for a longer period of $\sum_{\rm Q} > 2.6$ C/cm$^2$.

\subsection{$CE$ and Al-layer}

The drawback of introducting an Al prevention layer is a reduction of 
the photoelectron collection efficiency ($CE$) to $\sim 35\%$ from 
$\sim 60\%$ in the case of no Al-layer, as can be seen in Table \ref{T-2}. 
CT0790 and YA0071 are R3809U-50-11X circular-shaped PMT's, 
used in a previous study, but they were made especially for this research.
The CT0790 has an Al-layer between the photocathode and 1st-MCP, and 
the YA0071 does not. 

In order to optimize the Al-layer positioning, two different SL10's of 
YJ-versions were prepared with a $4\times 4$ anode configuration. 
The YJ0006 equips the Al-layer between the photocathode and the 1st MCP, 
and the YJ0011 has it between the 1st and 2nd MCP-layers. 
In the latter case, 1 mm of distance between 
the MCP's was prepared, and a HV was applied between them. 

The YJ0006 exhibits $CE$ = 36\%, just as the CT0790 does, 
while the YJ0011 results in $CE$ = 60\%, the maximum efficiency, which 
is equivalent to the MCP aperture of SL10 in the case of no Al-layer. 

Based on this study, all SL10's manufactured ever since are equipped 
with an Al-layer between the MCP-layers. 

\begin{table*}[bth]
\begin{center}
\caption{\small Al-layer effect on $CE$ }
\label{T-2}
\vspace*{2 mm}
\small
\begin{tabular}{|c||c|c|c|c|} \hline\hline
item & \multicolumn{2}{c|}{R3809U-50-11X}
     & \multicolumn{2}{c|}{SL10}  \\ \hline
PMT & CT0790 & YA0071 & YJ0006 & YJ0011  \\ \hline \hline
window & \multicolumn{2}{c|}{synthetic silica} 
       & \multicolumn{2}{c|}{borosilicate glass} \\ \hline
external size (mm$^3$) & \multicolumn{2}{c|}{45$^{\phi}\times70.2$} 
       & 27.5$\times$27.5$\times$12.1
       & 27.5$\times$27.5$\times$13.1 \\ \hline
effective area (mm$^2$) &  \multicolumn{2}{c|}{11$^{\phi}$} 
       & \multicolumn{2}{c|}{22$\times$22} \\ \hline
Al layer & cathode-1st MCP & No & cathode-1st MCP & 1st-2nd MCP \\ \hline
anode & \multicolumn{2}{c|}{single} 
& \multicolumn{2}{c|}{4$\times$4} \\ \hline
HV (v) & \multicolumn{2}{c|}{3,400} & 3,240 & 3,420 \\ \hline \hline
$QE$(@$\lambda=400$ nm) (\%) & 21 & 19 & 9$^{\sharp}$ & 23  \\ \hline
$CE$ (\%) & 37 & 65 & 36 & 60 \\
Gain ($\times 10^6$) & 1.5 & 1.9 & 0.41 & 1.1 \\\hline
TTS (ps) & 29 & 34 & 37 & 35 \\
dark counts (kcps) & 1.5 & 0.38 & 0.04 & 2.1 \\ \hline \hline
\end{tabular}
\normalsize
{\ }\\
$\sharp$: {\small Photocathode was not successfully made, so that 
$QE$ was low but $CE$ measurement would not be influenced. }
\end{center}
\end{table*}
\vspace*{10 mm}

\subsection{SL10 and lifetime $\tau_{\rm Q}$}

For our convenience, we define the lifetime, $\tau_{\rm Q}$, in terms of 
$\sum_{\rm Q}$, at which the $QE$ drops to 80\% ($\simeq -1$ dB) of 
the beginning. 
Within this $QE$ limit, the performance of the TOP counter could remain 
unaffected. 

\begin{figure*}[hbt]
\begin{center}
\resizebox{0.9\textwidth}{!}{\includegraphics{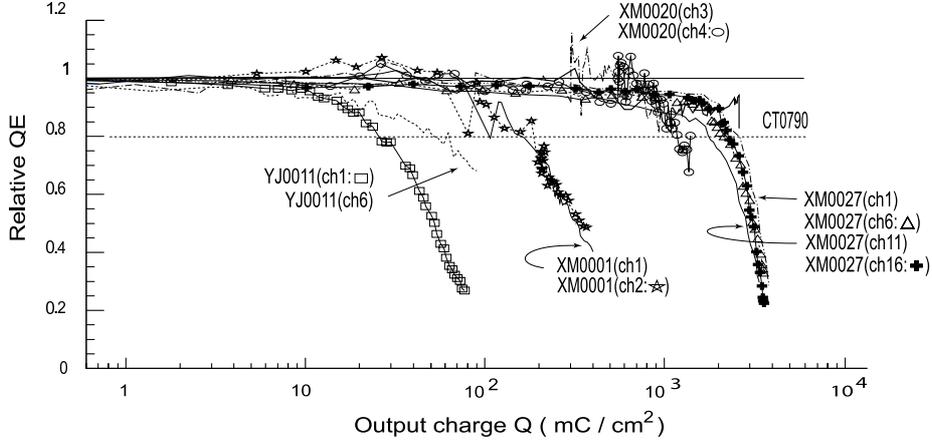}}
\caption{\small Relative $QE$ vs. $\sum_{\rm Q}$. 
Plotted are for R3809U-50-11X (CT0790) and SL10's
 (YJ0011, XM0001, XM0020 and XM0027). 
}
\label{F-1}
\end{center}
\end{figure*}

Figure \ref{F-1} shows the observed $QE$ vs. $\sum_{\rm Q}$ variation of 
the CT0790 and YJ0011 PMT's. 
The CT0790 exhibits a long life of $\tau_{\rm Q} > 3$ C/cm$^2$, as expected 
from previous measurements; however, it showed a small degradation that 
was not detected for the previous CT0790 up to 
$\sum_{\rm Q}\simeq 3$ C/cm$^2$.  

Concerning the YJ0011, one of the most corner located (1st) and 
one of the central (6th) 
anodes among the $4\times 4$ configuration (numbered in sequence from a corner) 
are plotted. 
In spite of furnishing the Al-layer, the YJ0011 exhibited a much shorter 
lifetime of $\tau_{\rm Q}=0.03-0.05$ C/cm$^2$, compared to the CT0790.  
The fact that the 1st-anode drops $QE$ faster than the 6th-anode is 
also observed (but not plotted) for the YJ0006. 

\begin{figure}[hbtp]
\begin{center}
\hspace*{-3 mm}
\resizebox{0.9\textwidth}{!}{\includegraphics{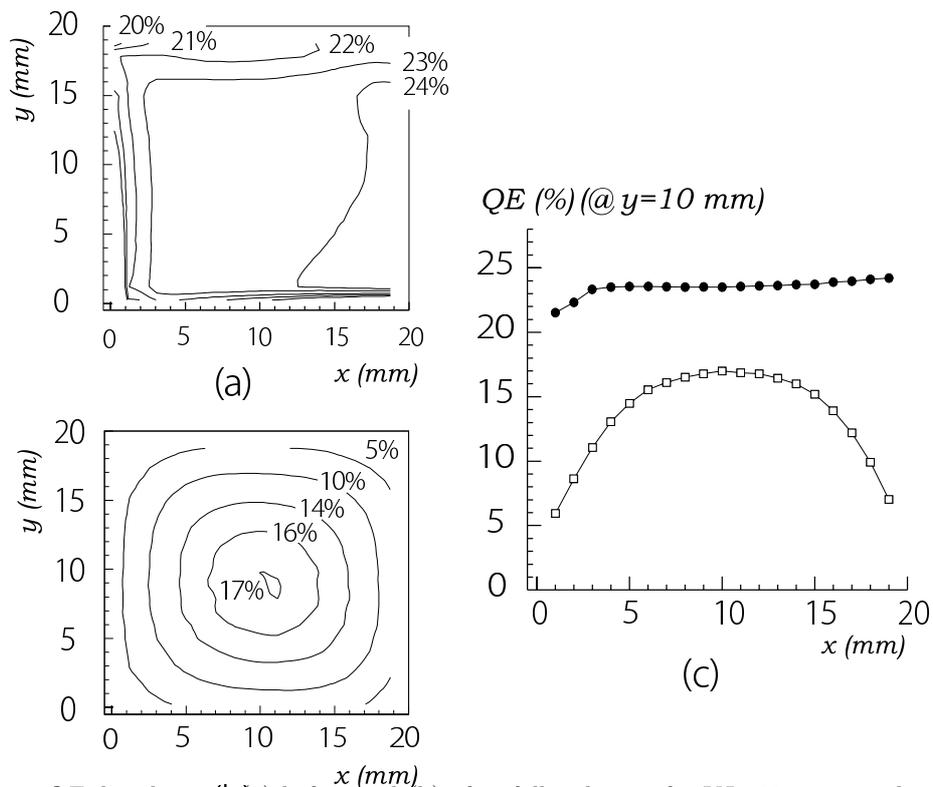}}
\vspace*{-10 mm}
\caption{\small $QE$ distributions (a) before and (b) after 
full radiation for YJ0011, measured using a monochromator system 
with $\lambda=400$ nm. (c) shows the $QE$ x-distributions before 
({\Large{$\bullet$}}) and after ($\Box$) at y=10 mm.}
\label{F-2}
\end{center}
\end{figure}

We measured the $QE$ distribution over the photocathode surface on YJ0011
by scanning with a slit of $1\times 1$ mm$^2$-size using 
a monochromator system at $\lambda=400$ nm;
Fig.\ref{F-2} shows the observed distributions before and after 
full irradiation of $\sum_{\rm Q}\simeq 0.065$ C/cm$^2$. 
The $QE$ distribution is quite homogeneous at the beginning, 
while it degrades, especially, further along the surrounding.
For instance, it varies from $QE\simeq 23-24$\% to $\simeq 10-17$\% at 
the vicinity of the center and from $20-24$\% to $\sim 5$\% along 
the sides. 

For the CT0790, the Al-layer essentially functions to protect 
the photocathode, but this is not the case for the SL10. 
There might exist another source(s) of deterioration. 

\subsection{Inner structure and residual gases}

\begin{center}
\begin{figure}[hbt]
\hspace{-3 mm}\resizebox{0.9\textwidth}{!}
{\includegraphics{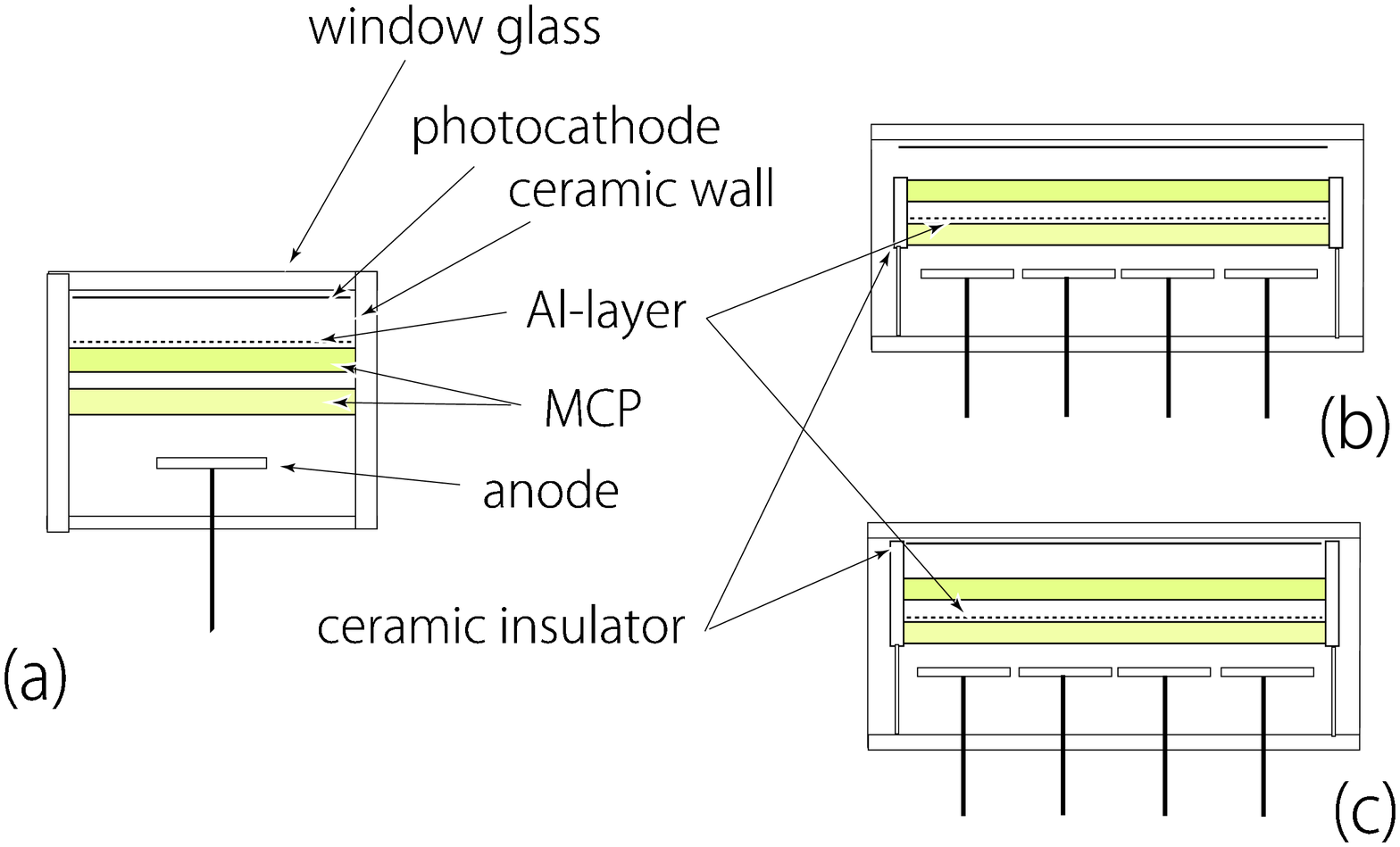}}
\caption{\small Schematic drawing of inner structure of PMT's. 
(a) CT0790, (b) YJ0011 and (c) XM-versions. 
Arrows for common items are omitted for (b) and (c). }
\label{F-3}
\end{figure}
\end{center}

Fig.\ref{F-3} illustrates the inner structure of two kinds of PMT's. 
The CT0790 holds MCP-layers adhered to a cylindrical ceramic container. 
On the other hand, the container of the SL10 is made of a cubic shape of 
metal in order to have sufficient mechanical strength to sustain 
a wider effective area, and its MCP-layers are held with different 
supporting parts for YJ0011. 

We speculate that some neutral molecular residual gases exist, bounced from 
the MCP's in the multiplication process of the secondary electrons, 
and whose production rate is proportional to the $\sum_{\rm Q}$, just the 
same as the production rate of the positive-ions, which
will damage the photocathode. 
Even so, it is expected that the residual gases are prevented by 
the Al-layer, just as the positive-ions are. 
It is inferred that it might be for the gases passing through the MCP 
channels to the photocathode, but it might not be for the gases making 
a detour through the side-ways between the MCP's and the metallic wall. 
As discussed later, gases such as carbon dioxide and monoxide, 
oxygen, and water, which can be the main ingredients of the residual gases,
are reported to affect the work function, $\phi$, of the photocathodes. 

Our research has since then focused on two subjects.
One is to confirm the above conjecture concerning neutral residual
gases; the other is to find measures to cope with their suppressions 
or removal. 

\section{Suppression of Gases} 

\subsection{Sealing photocathode from gases} 

We have manufactured various different SL10's, named as JT-versions 
(its structure is essentially the same as that of YJ0011, as listed 
in Table \ref{T-1} and \ref{T-2}, but with 4 anode-channels), 
each of which replaces some elements that attempt to suppress 
the residual gases.
Among many trails, for examples, 
MCP's are replaced with ones of different material-property;
MCP's are electron-scrubbed to further remove the residual gases; 
getters are furnished close to the photocathode to adsorb out-gases,   
and so on.  
No improvement in the lifetime has been obtained for any of the PMT's: 
$\tau_{\rm Q} < 0.1$ C/cm$^2$. 

Next, we employed ceramic elements, on the one hand, to support MCP-layers 
(Fig.\ref{F-3}(b) and (c)) 
and, on the other hand, to isolate the photocathode from the gap-space 
between the MCP's and metal-walls where the gases would be supposed 
to pass through (Fig.\ref{F-3}(c)). 
These SL10's are named ``XM-version'', and the insulation becomes 
tighter with the XM-version number. 
The lifetime now extends to $\tau_{\rm Q}\sim 0.1$ C/cm$^2$ by XM0001. 
We have found after many examinations that the tighter is the ceramic isolation,
the longer is the lifetime attained. 

Supposing that the MCP's are the source of residual gases outbreaking, 
MCP's are further treated to the highest degree of cleanliness, as 
mentioned above, in addition to furnishing 
the Al prevention layer between the MCP's. 
The SL10 has been developed with a succession of its manufacturing, and 
the longest lifetime now reaches $\tau_{\rm Q}=2- 3$ C/cm$^2$ by 
XM0027. 

\subsection{Gases vs. ion-feedback}

\begin{center}
\begin{figure}[H]
\resizebox{0.9\textwidth}{!}{\includegraphics{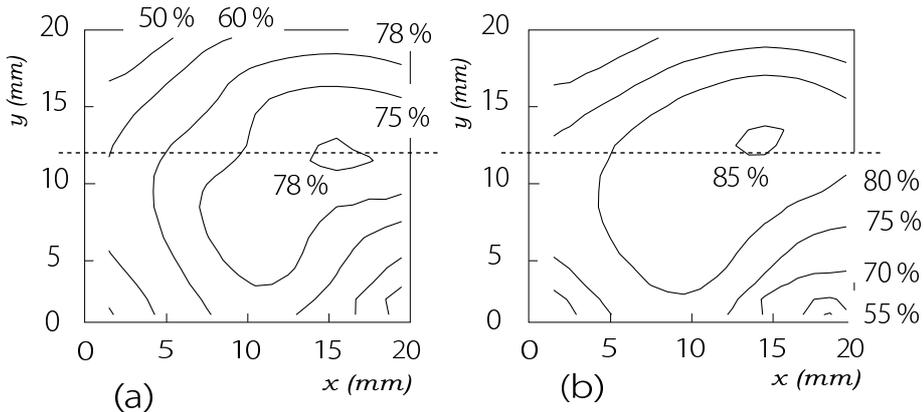}}
\vspace*{-5 mm}
\hspace*{-5 mm}
\caption{\small Contours of $QE$ drops for XM0001, measured with 
$\lambda=400$ nm. 
The integrated output charge, $\sum_{\rm Q}$, over the irradiation period 
is (a) $\simeq 0.2$ C/cm$^2$ and (b) $\simeq 0.16$ C/cm$^2$.
For (b) half of the photocathode surface ($y\geq 12$ mm), corresponding to 
the 3rd and 4th anodes, is shielded from the light. }
\label{F-4}
\end{figure}
\end{center}

In developing the XM-version, at an early stage we attained
a lifetime of $\tau_{\rm Q}=0.1-1$ C/cm$^2$.
The XM0001, having 4 anodes, is such a PMT (see, Fig.\ref{F-1}). 
Their $QE$'s of all 4-anodes drop to 70-80\% after irradiation 
of $\sum_{\rm Q}\simeq 0.2$ C/cm$^2$ (see, Fig.\ref{F-4}(a)); 
since then, half of its photocathode-surface ($y\geq 12$ mm), 
corresponding to the 3rd and 4th anode-channels, is shielded 
from irradiating light. 
If the $QE$ drop can be attributed, regardless of the Al-layer 
presence, to positive-ions or something else produced in 
the multiplication process and feedback through the MCP channels, 
the degradation over the shielded surface will stop. 
We, of course, do not observe the output signals from these anodes 
at the rates of the LED triggered, but do so from the 1st and 2nd anodes. 

Observation for a consecutive irradiation of $\sum_{\rm Q}\simeq 0.16$ 
C/cm$^2$ results in a quite similar $QE$ degradation, as can be seen in 
Fig.\ref{F-4}(b), with the former one. 
Table \ref{T-3} also lists the $QE$ drops for 4 anodes in the cases 
with and without half shielding. 
The $QE$ drops $\sim 20$\% 
more in the surrounding than that in the center area. 
\vspace*{-10 mm}
\begin{table}[h]
\caption{\small $QE$ drops for 4-anodes of XM0001 in cases of 
the photocathode surface, fully opened and half (3rd and 4th) shielded, 
respectively, after irradiation of $\sum_{\rm Q}\simeq 0.2$ C/cm$^2$ and 
$\simeq 0.16$ C/cm$^2$.  Measurements were made with $\lambda=400$ nm, and  
the observed $QE$ error was evaluated to be $\pm$1\%. }
\label{T-3}
\small
\begin{center}
\begin{tabular}{|l|cc|cc|} \hline\hline
anode channel & 1st & 2nd & 3rd & 4th \\ \hline\hline 
Full-opened (\%) & 68 & 71 & 71 & 63 \\ \hline
Half-shielded (\%) & 71 & 83 & 85 & 76 \\ \hline \hline
\end{tabular}
\end{center}
\end{table}

This fact supports our speculation that the residual gases taking 
the roundabout route, not directly through the MCP channels, to 
the photocathode would cause $QE$ deterioration. 

\section{SL10 with $\tau_{Q}\sim 2-3$ C/cm$^2$}

Along with steady progress by improving the individual elements and 
mechanical structure, we have obtained insulation-tight SL10's, 
the XM0020 and XM0027 with 4 and 4$\times$4 anodes, respectively, 
which attain $\tau_{\rm Q}\simeq 1$ and $2-3$ C/cm$^2$, 
as can be seen in Fig.~\ref{F-1}. 
The lifetime of the XM0027 is the longest one ever achieved, 
which corresponds to $12-19$ equivalent-years at the planed Super-KEKB / 
Belle-II. 

\begin{center}
\begin{figure}[H]
\vspace*{-5 mm}
\hspace*{-5 mm}
\resizebox{0.9\textwidth}{!}{\includegraphics{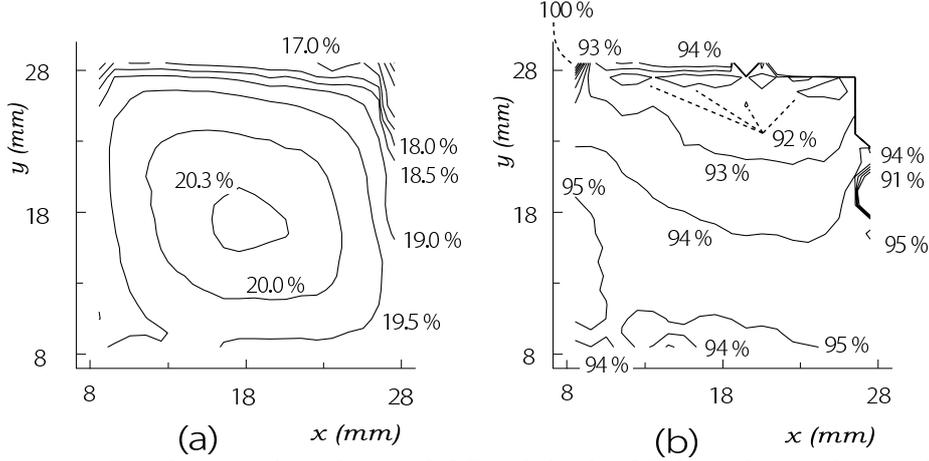}}
\vspace*{-5 mm}
\caption{\small Distributions of (a) the initial $QE$ and (b) the $QE$ 
drop after irradiation of $\sum_{\rm Q}\simeq 0.36$ C/cm$^2$ for 
XM0020, measured with $\lambda=400$ nm. }
\label{F-5}
\end{figure}
\end{center}

Homogeneous photocathode manufacturing on the square-shaped 
window has been established.
No strong degradation along the sides has been observed for these SL10's, 
as shown in Fig.~\ref{F-5}. 

\begin{center}
\begin{figure}[H]
\resizebox{0.8\textwidth}{0.6\textwidth}
{\includegraphics{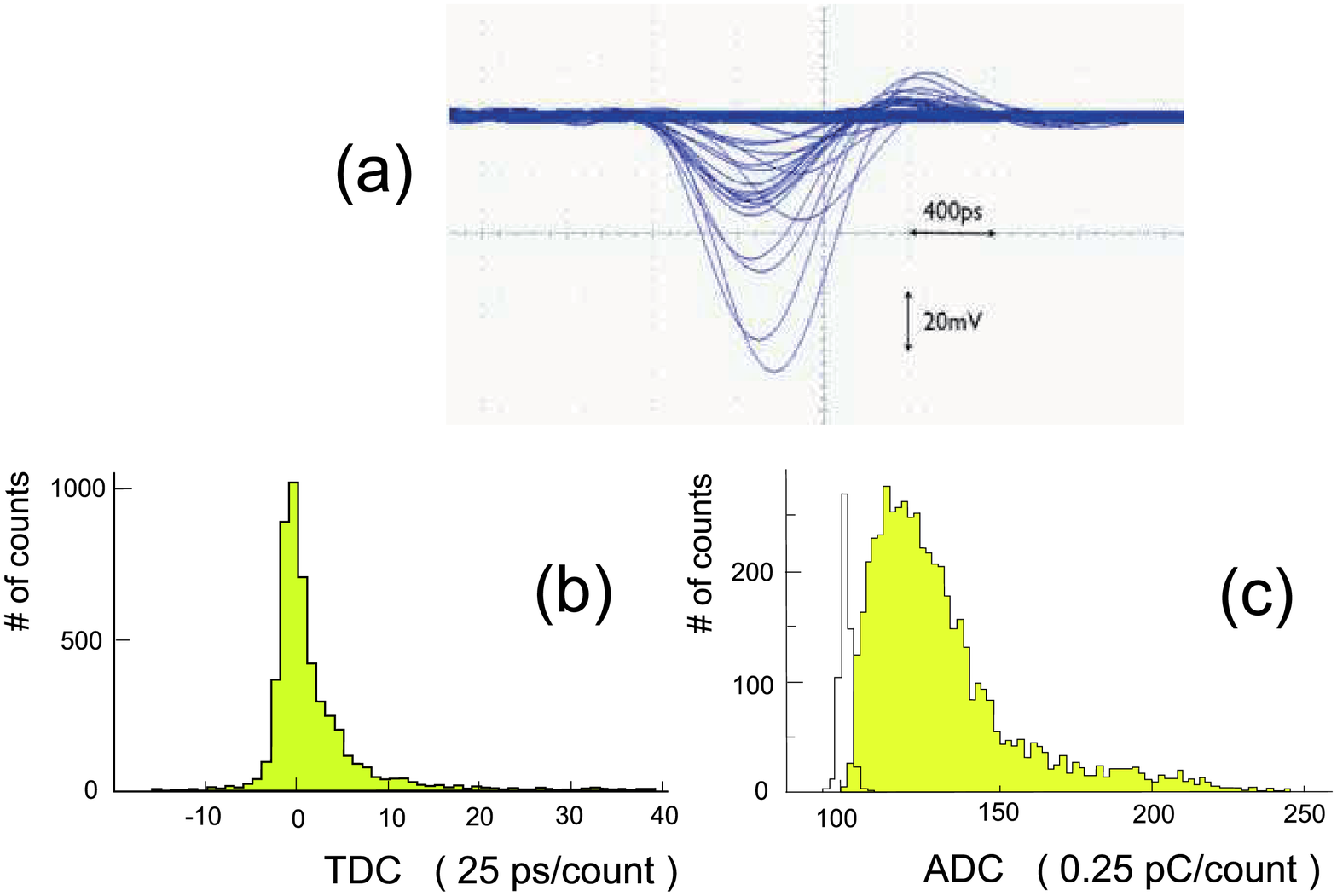}}
\resizebox{0.7\textwidth}{!}{\includegraphics{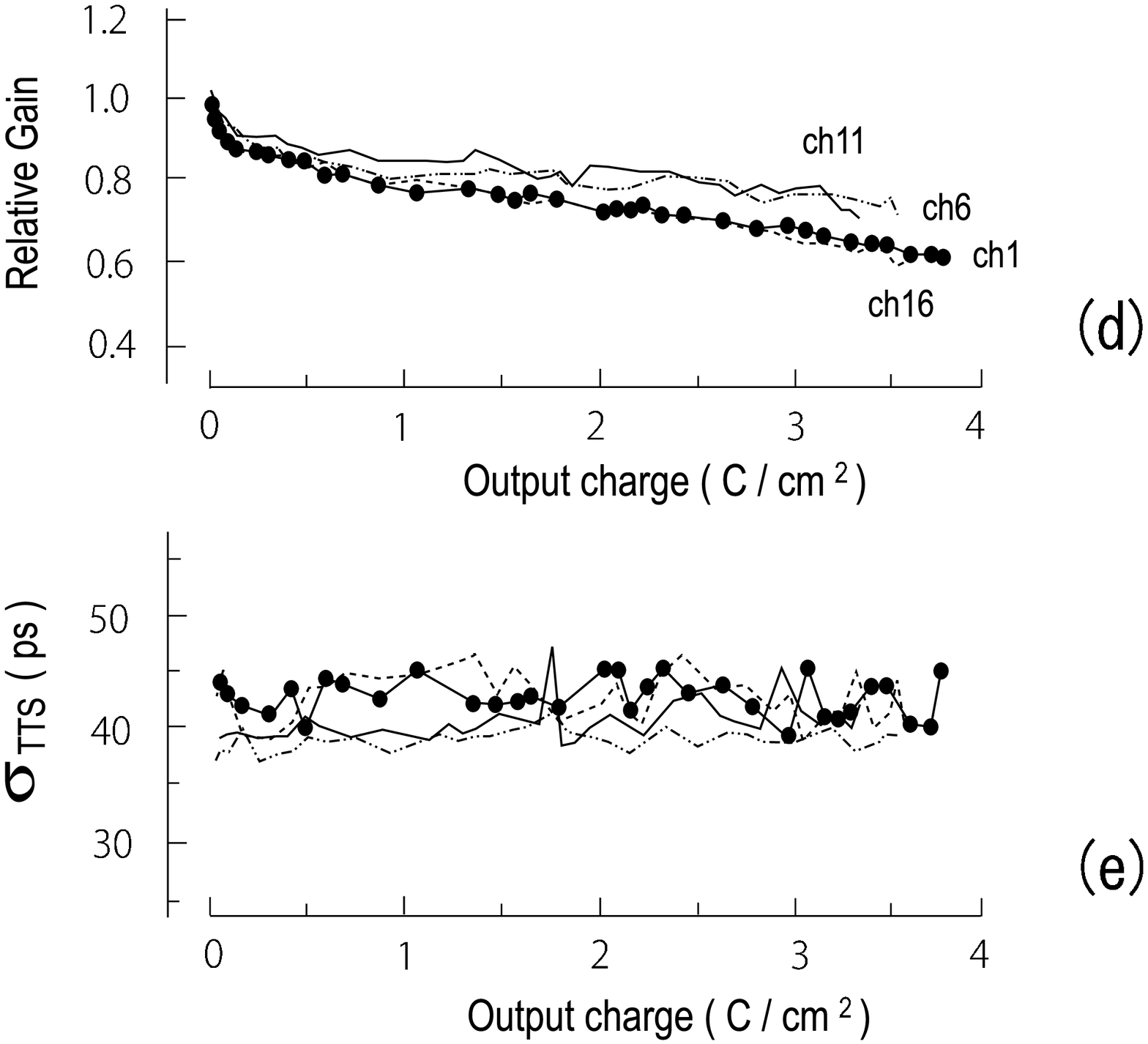}}
\caption{\small Performance of XM0020 and XM0027. 
(a) output signals, (b) TDC and (c) ADC distributions for XM0020, 
measured with HV=3,400 kV. 
(d) and (e) are the relative gain variation and the time resolution 
for single photons for XM0027. }
\label{F-6}
\end{figure}
\end{center}

Figure~\ref{F-6} shows their performance: 
the output signals for single photons, the ADC and TDC histograms 
for the XM0020, and 
the relative gain variation and the time resolutions, $\sigma_{\rm TTS}$,
for the XM0027 during the irradiation. 
The gain is $G\sim (1-3)\times 10^6$ in the beginning, but it 
linearly drops with $\sum_{\rm Q}$. For instance, it is 
reduced to 60\% at $\sum_{\rm Q}=3.8$ C/cm$^2$ for the XM0027 
(HV=3.8 kV).  
The $\sigma_{\rm TTS}$ value is stable, and exhibits a similar value 
between the two XM's; 
it varies within a range of $41\pm 4$ (ps) for the XM0027 
and within $46\pm 4$ (ps) for the XM0020. 

Dark counts of the XM0020 (HV=3.8 kV) are 
$\sim 18$ kHz at the 2nd anodes, 
but $\sim 1$ kHz at the other 3 anodes in the beginning, 
while they are $\sim 3$ kHz, but $\sim 10$ Hz, respectively, 
after the irradiation period. 
A similar situation also occurs for the XM0027.
They are 16, 36, 160 and 330 Hz at the 1st, 2nd, 3rd and 4th 
anodes in the beginning, and 0.2, 3, 4 and 17 Hz after 
irradiation. 

\section{Discussions and Results}

\subsection{Effect on the work function}

The photoelectron emission consists of four consecutive processes.
The absorption of a photon ($E=h\nu$) by the cathode material, 
the excitation of a valence electron to the conduction band, 
transfer of the excited electron to the cathode-surface, and 
the emission of an electron in a vacuum \cite{Spicer}. 
$QE$ is a product of the probabilities of these processes, and it is cut-off 
at a shorter wavelength of $\lambda\sim 300$ nm for borosilicate glass, 
fixed by the transmittance index of a window material, and 
at a longer wavelength of $\lambda\sim 900$ nm, settled by the work function 
($\phi\simeq 1.4$ eV of the multialkali ${\rm Na_2KSb(Cs)}$ photocathode). 

\begin{center}
\begin{figure}[H]
\resizebox{0.9\textwidth}{!}{\includegraphics{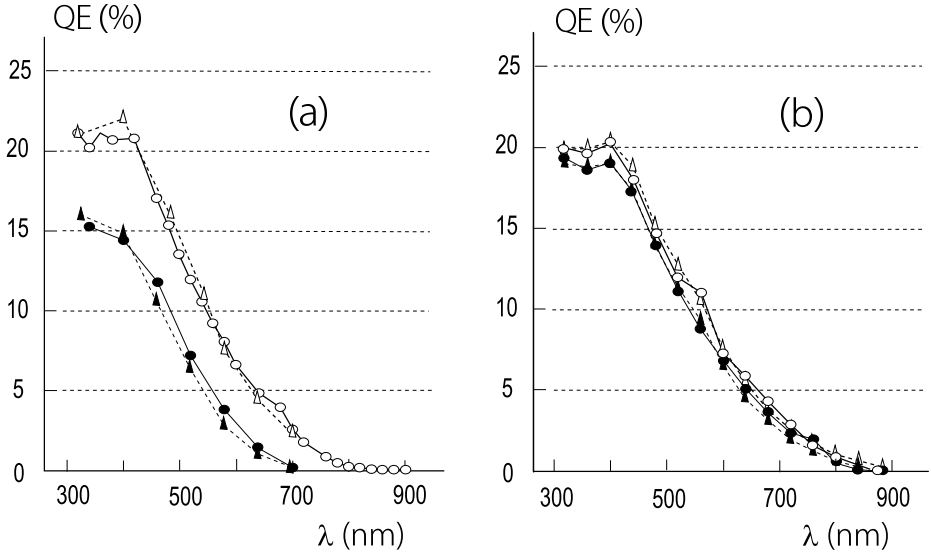}}
\caption{\small QE($\lambda$) spectrum changes:
(a) XM0001 with $\sum_{\rm Q}=0.2$ C/cm$^2$ and (b) XM0020 with 
$\sum_{\rm Q}=0.36$ C/cm$^2$. 
${\circ}$ and ${\triangle}$ are for the 1st and 2nd anodes 
at the beginning, and ${\bullet}$ and ${\blacktriangle}$ 
after irradiation.}
\label{F-7}
\end{figure}
\end{center}

As can be seen in Fig.\ref{F-7}(a), an irradiation of $\sum_{\rm Q}\simeq 0.2$ 
C/cm$^2$ on the XM0001 induces a shift of the longer wavelength-cutoff.
The $\phi$ value changes from 1.4 eV ($\lambda\simeq 900$ nm) to 
$\simeq 1.7$ eV ($\simeq 700$ nm). 
The XM0020, on the other hand, keeps $QE$ stable, even at 
$\sum_{\rm Q}\simeq 0.36$ C/cm$^2$ (see, Fig.\ref{F-7}(b)). 

\subsection{Cause of $QE$ degradation}

Let us itemize, without any prejudice, what we found concerning the cause: 
\begin{description}
\item[(1)]
Furnishing an Al-layer makes it possible to extend the lifetime,
$\tau_{\rm Q}$, by about two orders of magnitude, or more than 
that without the layer for the R3809U-50-11X. 
The Al-layer cuts across the inner structure between the photocathode 
and the 1st MCP, and it eliminates the after-pulse signals from the anodes 
\cite{Kishi}.
\item[(2)]
For SL10, while the Al-layer also splits the inner structure into two
(but between the MCP-layers), there still remains a gap-space between 
the MCP's and the container. 
Unless this gap-space is covered, 
an extension of the lifetime cannot be accomplished. 
\item[(3)]
Without covering the gap-space, the photocathode reduces its $QE$ 
over the whole area; the reduction is strong, especially along the sides 
of the container. 
Although half of the photocathode-surface is shaded from irradiating photons, 
$QE$ degradation occurs at the shaded area, just as 
the un-shaded region is affected. 
Covering the gap space extends the $QE$ lifetime to as long as 
$\tau_{\rm Q}=(2-3)$ C/cm$^2$. 
\item[(4)]
High degrees of cleaning and degassing of MCP-layers to reduce 
residual gases helps to improve the lifetimes. 
\end{description}

We previously conjectured positive-ion feedbacks to be the cause 
of $QE$ degradation. 
However, from the above-observed facts we now presume that 
the previous conjecture is not a unique explanation.
Residual gases are also the cause. 
Moreover, it is possibly the main reason. 

\subsection{Gases and photocathode's molecular structures}

We so far speculate that neutral residual gases bounced from 
the MCP-layers are the cause of $QE$ degradation. 
There are several reports on the effect of some gases on photocathodes.
T. Wada {\em et. al.} \cite{Wada} indicated that CO$_2$, CO and H$_2$O gases 
influence the $QE$ of (NEA) GaAs photocathodes, among which 
CO$_2$ molecules seem to have a serious influence; also
P. Michelato {\em et. al.} \cite{Miche} have reported that 
CO$_2$ and O$_2$ gases produce a rapid $QE$ drop for ${\rm K Cs Sb}$ 
bi-alkali photocathodes, 
but CH$_4$ and CO gases are sufficiently harmless. 
They measured the $QE$ change as a function of the exposure time duration 
under certain gas pressures; however, we do not have any information about 
the pressure and residual gases of our PMT's. 

Our brief interpretation of the data by T. Wada {\em et. al.} \cite{Wada} 
points out that the CO$_2$ or O$_2$ gases could yield an appreciable 
$QE$ degradation when the photocathode is exposed to the gas in terms of 
the product of its pressure and exposure duration of 
${\cal O}(\geq 10^{-10})$ (Torr$\times$h). 
By applying this criterion for our SL10, we estimate the gas pressure, $p$, 
when it arrives at its lifetime, $\tau_{\rm Q}$. 
The residual gases in our case are accumulated with time, 
so that the resulting effect would appear to be proportional to 
$\sum_{\rm Q}\times t$, where $t$ is the full time-length since 
the beginning of irradiation.  
An order evaluation suggests $p\sim {\cal O}(10^{-13})$ (Torr) 
for the XM0027. 

C. Ghosh and B. P. Varma \cite{Ghosh} studied alkali antimonide 
photocathodes, and measured the work functions to be $\phi\sim 1.4$ eV 
for ${\rm Na_2KSb(Cs)}$ and 1.8 eV ($\lambda\simeq 680$ nm), 
1.9 eV ($\lambda\simeq 650$ nm) and 2.2 eV ($\lambda\simeq 560$ nm) 
for ${\rm Na_2KSb}$, ${\rm K_2SbCs}$ and ${\rm K_3Sb}$ structures, 
respectively. 
Accordingly, the variations of our $QE$($\lambda$) and 
$\lambda$-cutoff before and after a large irradiation might 
indicate a change in the photocathode structure from that of conventional 
${\rm Na_2KSb(Cs)}$ to ${\rm Na_2KSb}$. 
They have the same band-gap of 1.1 eV, 
but their work functions are different by $\sim$ 0.4 eV. 

T. Guo and H. Gao \cite{Guo} have discussed rules concerning cesium and oxygen 
on the energy levels of several kinds of photocathodes.
An excess of oxygen will reside on top of the Cs activation-layer, 
resulting in a ${\rm Na_2KSb-Cs-O}$ structure rather than 
the nominal ${\rm Na_2KSb-O-Cs}$ structure, and 
then increase $\phi$ and decrease $QE$. 

\subsection{Ideal measure} 

It is acceptable to think that the Al-layer, made by evaporating 
Aluminum to a few ten's of $\AA$-thick, could substantially prevent 
positive ions and residual gases from passing through. 
Our R3809U-50-11X still does not show any obvious $QE$-drop up to 
$\sum_{\rm Q}\simeq 3$ C/cm$^2$. 

Besides the Al-layer, 
although the insulation of the gap-space from the photocathode is 
done in the case of SL10, it is not air-tight, but rather black-outing, 
so that the gases could leak into the photocathode domain, and 
might lead to a $QE$ drop, as observed.  

Accordingly, the most ideal way would be to realize the inner 
structure to be the same as the R3809U-50-11X.
Both mechanical and electric considerations must be required. 
If it can be materialized, the lifetime could be extended to be much 
longer than that of XM0027. 

\subsection{Aging}

It has been found that some PMT's exhibit rapid decreases of $QE$ and $G$ 
in the first $\sum_{\rm Q}\sim 0.1$ C/cm$^2$. 
It drops by $5-10$\%, which can be seen in Fig.\ref{F-6}(d) on $G$. 
The gain, $G$, also slowly decreases linearly with $\sum_{\rm Q}$, and 
a similar behavior occurs concerning the rate of the dark counts, mentioned 
in the previous section for XM0020 and XM0027. 

All of those effects might be ascribed to aging phenomena of the PMT. 
Its effect is most remarkable for the dark counts.
Their rates decrease by about two orders of magnitude during irradiation. 

\subsection{Results}

As a photo-device for the TOP counter, planned for the second 
generation of the high-luminosity B-factory Super-KEKB/Belle-II, 
we have developed an MCP-PMT, SL10, having considerable endurance 
under high rates of counting for a long experimental period. 
It is found that in addition to positive ion-feedbacks, residual gases 
had substantial influences on the lifetime of the PMT. 
The $QE$ variation with the integrated amounts of the output charge, 
$\sum_{\rm Q}$, was measured, and the $QE$($\lambda$) spectra before 
and after irradiation were examined for about 30 different versions 
of SL10's. 

Both the furnishments of the Al-layer and the ceramic-insulation 
largely suppress the ions and gases from damaging the photocathode, 
and then resulting in a long lifetime 
of $\tau_{\rm Q}\simeq 2-3$ C/cm$^2$, which is 
equivalent to well more than 10 years of operation at 
the supposed Super-KEKB/Belle-II. 

The resulting SL10 with $4\times 4$ anodes, XM0027, exhibits in
a satisfactory performance with $\sigma_{\rm TTS}\simeq 40$ ps, 
QE($\lambda=400$ nm)$\simeq$ 20\%, CE=60\%, 
G=$(1-3)\times 10^6$ and dark counts of $20-300$ Hz.

{\ }\\
\noindent
{\bf Acknowledgments}\\

This work is supported by a Grant-in-Aid for Science Research 
in a Priority Area (``New Development of Flavor Physics'') from 
the Ministry of Education, Culture, Sports, Science and Technology of 
Japan, and from the Japan Society for the Promotion of Science for 
Creative Scientific Research (``Evolution of Tau-lepton Physics''). 
We acknowledge support from the Tau-Lepton Physics Research Center of 
Nagoya University. 
%



\begin{thebibliography}{99}
\bibitem{ZERO}
For the lifetime evaluation we suppose the detector condition with
a luminosity of $2\times 10^{35}$/cm$^2$/s in the ref.\cite{Kishi}, 
while new condition with $4\times 10^{35}$/cm$^2$/s in this time. 
Also, typical PMT performance is set as $CE$=50\% and $G=2\times 10^6$ 
previously but 60\% and $1\times 10^6$ here while keeping $QE$ as 20\%. 
We also take account of actual operation time per year.
Therefore, the conversion rate from $Q$ to the experimental equivalent-year 
is different between two cases.
\end{thebibliography}
\end{document}